\documentclass[twocolumn]{aastex62}

\submitjournal{ApJ}
\usepackage[all]{hypcap} 
\usepackage{tabularx, graphicx}
\usepackage{comment}
\usepackage{bm}
\usepackage{amsmath}
\usepackage{color}

\newcommand{\LENSTOOL}{{\tt{Lenstool}}}
\newcommand{\Lenstool}{{\tt{Lenstool}}}

\newcommand{\hst}{{\it HST}}
\newcommand{\HST}{{\it HST}}
\newcommand{\Hubble}{{\it Hubble}}

\newcommand{\dl}{D_\mathrm{L}}
\newcommand{\zs}{z_\mathrm{S}}
\newcommand{\zl}{z_\mathrm{L}}

\newcommand{\eER}{\theta_{\rm{arcs}}} 
\newcommand{\ER}{\theta_{\rm{E}}} 
\newcommand{\RSHM}{e\theta_{\rm{E,SHM}}}
\newcommand{\ML}{M(<\theta_{\rm E})} 
\newcommand{\cML}{M_{corr}(<\eER)}
\newcommand{\MSHM}{M_{\rm SHM}}
\newcommand{\MDLM}{M_{\rm DLM}}

\newcommand{\RcMLAllScatter}{$10.9\%$}

\newcommand{\RcMLAllLinScatter}{$12.1\%$}

\newcommand{\RMSHMAllScatter}{$8.5\%$}
\newcommand{\RMSHMAllBias}{$0.9\%$}

\newcommand{\RMPSHMAllScatter}{$3.3\%$}
\newcommand{\RMPSHMAllBias}{$0.3\%$}

\newcommand{\cMLAllScatter}{$18.1\%$}
\newcommand{\cMLAllBias}{$-7.1\%$}

\newcommand{\cMLSmallERScatter}{$14.4\%$}
\newcommand{\cMLSmallERBias}{$-4.3\%$}

\newcommand{\MSHMAllScatter}{$12.4\%$}
\newcommand{\MSHMAllBias}{$2.4\%$}

\newcommand{\MPSHMAllScatter}{$8.2\%$}
\newcommand{\MPSHMAllBias}{$1.0\%$}

\newcommand{\DLMstatsig}{$1.1\%$}

\newcommand{\EcMLAllscatter}{$12.0\%$}
\newcommand{\EMSHMAllScatter}{$6.0\%$}

\begin{document}

\title{Core Mass Estimates in Strong Lensing Galaxy Clusters: a Comparison Between Masses Obtained from Detailed Lens Models, Single-Halo Lens Models, and Einstein Radii}

\author[0000-0002-7868-9827]{J. D. Remolina Gonz\'{a}lez}
\affiliation{Department of Astronomy, University of Michigan, 1085 S. University Ave, Ann Arbor, MI 48109, USA}
\email{jremolin@umich.edu}

\author[0000-0002-7559-0864]{K. Sharon}
\affiliation{Department of Astronomy, University of Michigan, 1085 S. University Ave, Ann Arbor, MI 48109, USA}

\author[0000-0003-3266-2001]{G. Mahler}
\affil{Department of Astronomy, University of Michigan, 1085 S. University Ave, Ann Arbor, MI 48109, USA}
\affil{Centre for Extragalactic Astronomy, Department of Physics, Durham University, Durham DH1 3LE, UK}
\affil{Institute for Computational Cosmology, Durham University, South Road, Durham DH1 3LE, UK}

\author[0000-0001-8316-9482]{C. Fox}
\affil{Department of Astronomy, University of Michigan, 1085 S. University Ave, Ann Arbor, MI 48109, USA}

\author{C.A. Garcia Diaz}
\affil{Division of Physics, Engineering, Mathematics, and Computer Science, Delaware State University, 1200 N. Dupont Hwy, Dover, DE, United States}

\author[0000-0003-4470-1696]{K. Napier}
\affil{Department of Astronomy, University of Michigan, 1085 S. University Ave, Ann Arbor, MI 48109, USA}

\author[0000-0001-7665-5079]{L. E. Bleem}
\affil{Argonne National Laboratory, High-Energy Physics Division, Argonne, IL 60439}
\affil{Kavli Institute for Cosmological Physics, University of Chicago, 5640 South Ellis Avenue, Chicago, IL 60637, USA}

\author[0000-0003-1370-5010]{M. D. Gladders}
\affil{Department of Astronomy and Astrophysics, University of Chicago, 5640 South Ellis Avenue, Chicago, IL 60637, USA}
\affil{Kavli Institute for Cosmological Physics, University of Chicago, 5640 South Ellis Avenue, Chicago, IL 60637, USA}

\author[0000-0001-6800-7389]{N. Li}
\affil{CAS, Key Laboratory of Space Astronomy and Technology, National Astronomical Observatories, A20 Datun Road, Chaoyang District, Beijing 100012, People’s Republic of China}
\affil{School of Physics and Astronomy, Nottingham University, University Park, Nottingham NG7 2RD, UK}

\author[0000-0003-3791-2647]{A. Niemiec}
\affil{Department of Astronomy, University of Michigan, 1085 S. University Ave, Ann Arbor, MI 48109, USA}
\affil{Centre for Extragalactic Astronomy, Department of Physics, Durham University, Durham DH1 3LE, UK}
\affil{Institute for Computational Cosmology, Durham University, South Road, Durham DH1 3LE, UK}



\begin{abstract}

The core mass of galaxy clusters is both an important anchor of the radial mass distribution profile and probe of structure formation. With thousands of strong lensing galaxy clusters being discovered by current and upcoming surveys, timely, efficient, and accurate core mass estimates are needed. We assess the results of two efficient methods to estimate the core mass of strong lensing clusters: the mass enclosed by the Einstein radius ($\ML$, where $\theta_{\rm E}$ is approximated from arc positions; \citealt{Remolina:20}), and single-halo lens model ($\MSHM$; \citealt{Remolina:21}), against measurements from publicly available detailed lens models ($\MDLM$) of the same clusters. We use data from the Sloan Giant Arc Survey, the Reionization Lensing Cluster Survey, the \Hubble\ Frontier Fields, and the Cluster Lensing and Supernova Survey with \Hubble. We find a scatter of \cMLAllScatter\ (\MPSHMAllScatter) with a bias of \cMLAllBias\ (\MPSHMAllBias) between $\cML$ ($\MSHM$) and $\MDLM$. Last, we compare the statistical uncertainties measured in this work to those from simulations. This work demonstrates the successful application of these methods to observational data. As the effort to efficiently model the mass distribution of strong lensing galaxy clusters continues, we need fast, reliable methods to advance the field. 

\end{abstract}



\section{Introduction} 
\label{sec:intro}

Galaxy clusters are harbored at the knots of the cosmic web and trace the large-scale structure of the universe, making them ideal cosmic laboratories (see reviews by \citealt{Allen:11} and \citealt{Mantz:14}). The galaxy cluster mass function connects the underlying cosmology and the observational properties of galaxy clusters (e.g., \citealt{Evrard:02,Pratt:19,Bocquet:20}). Additional predictions from cosmological simulations include the radial mass distribution of dark matter halos (e.g., \citealt{Duffy:08,Meneghetti:14,Child:18}), which can be directly tested against observations via the concentration measurement (e.g., \citealt{Oguri:12,Merten:15}). An accurate account of the cluster mass distribution requires mass estimates that are sensitive at the cores and at the outskirts of the galaxy cluster. Crucial to all cluster-based cosmological studies are the sample size, selection function, and good understanding of the systematic uncertainties of the mass estimates coming from observed astrophysical properties (e.g., \citealt{Evrard:02,Khedekar:13,Huterer:18,Bocquet:19}). 

One of the methods to measure the total (dark and baryonic) mass distribution of galaxy clusters is using gravitational lensing. Weak lensing (WL) measures the cluster mass at large cluster-centric radii, while strong lensing (SL) has the highest resolution at the core of the cluster where the SL evidence is present. The combination of the core mass estimates from SL and outskirts mass estimates from WL or other large scale mass proxies can constrain the mass distribution profile of a galaxy cluster, and measure its concentration (e.g., \citealt{Gralla:11,Oguri:12,Merten:15,Meneghetti:10}). Comparisons between the predicted and observed properties of SL galaxy clusters mass distribution have reported possible tension (e.g., \citealt{Broadhurst:08,Gonzalez:12,Meneghetti:13,Killedar:18}), however these studies have been limited by complicated selection function and small sample sizes.

Current and upcoming large surveys will discover thousands of SL clusters out to $z \sim 2$, using methods that span the wavelength spectrum. Some of these surveys include the South Pole Telescope (SPT; SPT-3G, \citealt{Benson:14}; SPT-SZ 2500 deg$^2$, \citealt{Bleem:15}), Vera Rubin Observatory Legacy Survey of Space and Time (LSST, \citealt{LSST:17}), and eROSITA \citep{Pillepich:18}. These large samples will require a method to timely, effectively, and accurately measure the core mass of SL clusters.  

Strong-lensing based mass measurements are typically based on detailed strong lensing models (e.g., \citealt{Kneib:11}). Detailed lens models for galaxy clusters with rich strong lensing evidence, such as the Frontier Fields clusters \citep{Lotz:17} but also less extraordinary clusters, allow for the high degree of complexity required to study substructure in the mass distribution of the cluster (e.g., \citealt{Ebeling:17, Mahler:18, Richard:20}). They necessitate extensive follow-up observations, computational resources, and multiple statistical assessments for the best model selection. However, more typical SL clusters have a small number of SL constraints, which limits the utility of detailed lens models (e.g., \citealt{Smith:05, Sharon:20}). 

The large sample sizes of SL clusters being discovered call for efficient methods to estimate the mass at the core of galaxy clusters. \citet{Remolina:20} and \citet{Remolina:21} evaluated two methods for efficiently estimating the mass within the core of SL clusters using the Outer Rim cosmological simulation. \citet{Remolina:20} evaluated the mass estimate derived from the equation of the Einstein radius of a circularly symmetric lens, and \citet{Remolina:21} assessed results from simplified single-halo lens models. The characterization of uncertainty and bias of these methods established them for the application to large samples of SL galaxy clusters as efficient and accurate galaxy cluster core mass estimators. The two simulation-calibrated methods take orders of magnitude less time and human intervention than detailed lens models. 

The goal of this paper is to test, in real observed clusters, how well these first- and second-order estimates of the core mass compare to detailed lens models. This paper is organized as follows. In \S\ref{sec:data}, we introduce the four strong lensing cluster samples used in our paper and describe our selection of the detailed lens models. In \S\ref{sec:lens_modeling}, we briefly describe the publicly available lensing algorithms used to compute the detailed lens models and summarize the single-halo lens models and Einstein radius methods used as efficient estimates of the mass at the core of galaxy clusters. In \S\ref{sec:methodology}, we describe the strong lensing constraints and selection of the brightest cluster galaxy (BCG), estimate an approximation of the Einstein radius ($\eER$) from the observed lensing constraints, compute the empirically-corrected enclosed mass using the Einstein radius equation, $\cML$, and compute the aperture mass measured utilizing the single-halo lens models that passed a quick visual inspection, $\MSHM$. In \S\ref{sec:results}, we measure the scatter and bias of $\cML$ and $\MSHM$ compared to the mass enclosed by the same aperture in the detailed lens model ($\MDLM$) and explore any possible difference due to the variety of lensing algorithms utilized to compute the detailed lens models. Last in \S\ref{sec:conclusion}, we present our conclusions and summarize the application of efficient methods to measure the core masses of galaxy clusters.

In our analysis, we adopt a flat $\Lambda$CDM cosmology: $\Omega_{\Lambda} = 0.7$, $\Omega_{M} = 0.3$, and $H_{0} = 70$ km s$^{-1}$ Mpc$^{-1}$. The large scale masses are reported in terms of M$_{\mathrm{\Delta c}}$, defined as the mass enclosed within a radius at which the average density is $\Delta$ times the critical density of the universe at the cluster redshift.


\section{Observational Data} 
\label{sec:data}

\begin{figure}
\includegraphics[width=0.5\textwidth]{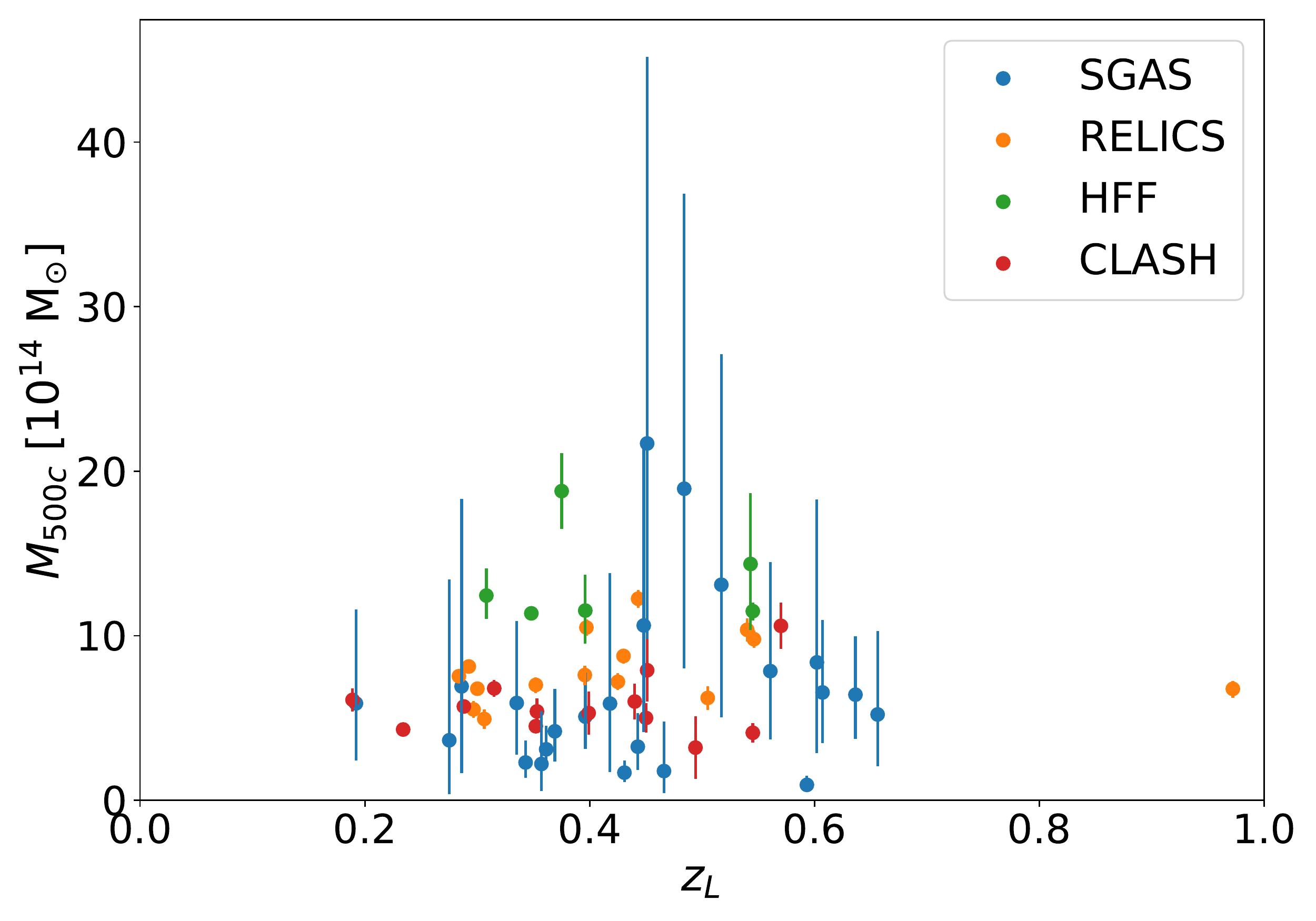}
\caption{Redshift-Mass ($\zl$ - $M_{500c}$) distribution of the strong lensing galaxy clusters used in our analysis.}
\label{fig:z_M_dist}
\end{figure}

For this work, we use the data from four well-established strong lensing surveys of clusters with different selections functions. First, the Sloan Giant Arcs Survey (SGAS\footnote{https://archive.stsci.edu/pub/hlsp/sgas/}; \citealt{Hennawi:08,Sharon:20}) which identified highly magnified lensed galaxies in the Sloan Digital Sky Survey (SDSS; \citealt{Abazajian:09,Blanton:17}). Second, the Cluster Lensing and Supernova Survey with \textit{Hubble} (CLASH\footnote{https://www.stsci.edu/~postman/CLASH/index.html}; \citealt{Postman:12}), designed to study the dark matter distribution in galaxy clusters, perform supernova searches, and detect and characterize high-redshift lensed galaxy clusters. Third, the \Hubble\ Frontier Fields Clusters (HFF\footnote{https://outerspace.stsci.edu/display/HPR/HST+Frontier+Fields}; \citealt{Lotz:17}) which are some of the best strong lensing clusters, taking advantage of deep imaging and extensive spectroscopic follow-up. Fourth, the Reionization Lensing Cluster Survey (RELICS\footnote{https://relics.stsci.edu/index.html}; \citealt{Coe:19}), designed primarily to find high-redshift ($z \sim 6 - 8$) lensed galaxy candidates. All four samples base their lensing analyses on multi-band \textit{Hubble Space Telescope} (\HST) imaging. From these samples of lensing galaxy clusters, we only include clusters with spectroscopically confirmed multiply imaged lensed galaxies. \autoref{fig:z_M_dist} shows the redshift-mass distribution of the galaxy clusters used in our analysis. The large scale masses, $M_{500c}$, are taken from \citet{Fox:21}, \citet{Merten:15}, and references therein.

\subsection{SGAS} 
\label{subsec:sgas}

Galaxy clusters where selected for the SGAS survey from the SDSS Data Release $7$ (SDSS-DR7; \citealt{Abazajian:09}) using the cluster red-sequence algorithm by \citet{Gladders:00}. Color images were created from imaging data in \textit{g}, \textit{r}, \textit{i}, and \textit{z} centered on the detected cluster. The images were visually inspected and scored according to the evidence of strong gravitational lensing. There has been extensive ground- and space-based imaging leading to a multi-wavelength picture of these clusters (e.g., 107 \textit{HST} orbits of the SGAS-\textit{HST}, GO13003 PI:Gladders; \citealt{Sharon:20}). Spectroscopic follow-up of all the primary strong lensing arcs is complete, and additional follow-up campaigns obtained redshifts of secondary arcs to improve the fidelity of the lens models (e.g., \citealt{Bayliss:11b, Johnson:17a, Sharon:20}, and references therein). Several high-impact targets out of this sample of highly magnified arcs have been studied in detail (e.g., \citealt{Koester:10, Bayliss:14, Sharon:17, Rigby:18}).

The SGAS clusters are unique due to the selection function employed to create the sample, as they were selected uniquely based on the identification of bright strong lensing features. This led to including some clusters with lower masses when compared to the other three samples of galaxy clusters. \citet{Sharon:20} published and released to the community detailed lens models for the $37$ SGAS clusters observed as part of \hst-GO13003. Out of these $37$ galaxy clusters with publicly available lens models \citep{Sharon:20}, we only use $31$ in this work. Three clusters (SDSS~J0004$-$0103, SDSS~J1002$+$2031, and SDSS~J1527$+$0652) are not included due to being poorly constrained (given a classification of C or lower; see \citet{Sharon:20} for more details). Two galaxy clusters (SDSS~J1156$+$1911 and SDSS~J1632$+$3500) lack any spectroscopically confirmed multiply imaged sources. One galaxy cluster (SDSS~J2243$-$0935), has one spectroscopically confirmed flat giant arc, located between two cluster cores, making it unsuitable for the methods used here. In \autoref{table:sl_table}, the list of the SGAS clusters is shown with their corresponding redshift, the right ascension (R.A.) and declination (Decl.) of the selected BCG, and the number of strongly lensed background sources with spectroscopic redshifts that are used as lens modeling constraints.

\subsection{CLASH}
\label{subsec:clash}

The Cluster Lensing And Supernova survey with \textit{Hubble} (CLASH; \citealt{Postman:12}) multi-cycle treasury project observed $25$ galaxy clusters for a total of $525$ \HST\ orbits over a period of nearly three years utilizing $16$ \HST\ filters. The main science goals included: studying the matter distribution of galaxy clusters, particularly the mass concentration (e.g., \citealt{Merten:15}); detecting supernovae (e.g., \citealt{Graur:14}); and detecting and characterizing high-redshift galaxies magnified by the galaxy cluster (e.g., \citealt{Coe:13}). From the $25$ galaxy clusters, $20$ are X-ray selected, dynamically relaxed (determined from their circularly symmetric X-ray surface brightness distribution), and massive clusters (X-ray temperatures T$_{\mathrm{x}} > 5$ keV). The majority of these clusters showed strong lensing evidence from ancillary data. The last five galaxy clusters were selected solely for being exceptional strong lenses. Four of the galaxy clusters (Abell\,S1063, MACS\,J0416.1$-$2403, MACS\,J0717.5$+$3745, and MACS\,J1149.5$+$2223) were later selected for the Hubble Frontier Fields (HFF; see \S\ref{subsec:hff}) and we only utilize the HFF lens models for these clusters. The community follow-up effort has resulted in the identification of many lensing constraints with measured spectroscopic redshifts for the $13$ galaxy clusters included in this work. Detailed lensing models by \citet{Zitrin:15} and \citet{Caminha:19} have been made publicly available. In \autoref{table:sl_table}, we list the CLASH galaxy clusters utilized in our analysis and their corresponding references.

\subsection{HFF}
\label{subsec:hff}

The \textit{Hubble} Frontier Fields (HFF; \citealt{Lotz:17}) project  observed six galaxy clusters and adjacent (``parallel'') fields using Director's discretionary time, obtaining extremely deep multi-band imaging ($140$ \textit{HST} orbits per cluster for a total of $840$ \textit{HST} orbits of Director's Discretionary Time) with the primary goal of studying the magnified background universe. The clusters were selected for their observability from space- (\textit{HST}, \textit{Spitzer}, and \textit{JWST}) and ground-based observatories, their lensing strength, and the availability of pre-existing ancillary data. These galaxy clusters have become some of the most studied galaxy clusters due to the community investment in extensive multi-wavelength imaging and spectroscopic follow-up, resulting in large numbers of strong lensing constraints identified and used in the detailed lens models (\citealt{Johnson:14, Zitrin:14, Diego:16, Jauzac:16, Limousin:16, Caminha:17, Karman:17, Kawamata:18, Mahler:18, Strait:18, Lagattuta:19, Sebesta:19, Vega-Ferrero:19, Raney:20a}; and references therein). The HFF program provides a unique opportunity to study the statistical and systematic uncertainties in the lensing outputs, due to the large number of diverse lensing algorithms that have computed detailed lens models of these clusters (e.g., \citealt{Meneghetti:17, Priewe:17, Remolina:18, Raney:20b} compare different aspects of the HFF lens models using different algorithms). In this work, we include the fourth version of the public lens models, which is the most recent release. The clusters and references to the models are listed in \autoref{table:sl_table}.

\subsection{RELICS} 
\label{subsec:relics}

The RELICS program selected $41$ galaxy clusters for shallow multi-band observation with \hst\ for the primary goal to deliver a large sample of high-redshift ($z \sim 6 - 8$) galaxies \citep{Salmon:18, Salmon:20, Mainali:20, Strait:20}. $21$ clusters where selected from a subsample of the most massive Planck clusters (using the Sunyaev-Zel\'dovich effect, \citealt{Sunyaev:70}, to estimate their mass; \citealt{Planck:16}). The other $20$ cluster were selected based on a prior identification as prominent strong lenses in available imaging data. The reasoning used for this selection is the expectation that the mass of the galaxy cluster relates to its potential to have a large lensing cross-section, leading to an increase in the chance to find high-redshift lensed sources. 

The selection function employed for assembling the list of RELICS clusters explores the high-mass parameter space. In addition, the wider and shallower imaging observing strategy (total of $188$ \textit{HST} orbits, GO 14096; PI Coe) is a clear example of the challenges confronted by lensing surveys where only the primary and some of the secondary arcs are readily identifiable, leading to a limited number of constraints available for the lens modeling analysis (\citealt{Acebron:18, Acebron:19, Acebron:20, Cerny:18, Cibirka:18, Paterno-Mahler:18,  Mahler:19} and references therein). From the $41$ galaxy clusters observed, $34$ have publicly available detailed lens models and only $17$ have publicly available spectroscopically confirmed multiple imaged sources. Following \citet{Fox:21}, we inspect the unpublished detailed lens models and include in our analysis only models whose predicted lensed images are within $1\farcs5$ of the observed lensing evidence, and do not produce critical curves or masses that are not justified by the lensing constraints. In \autoref{table:sl_table}, we present the list of the RELICS clusters used in our analysis with their corresponding lens redshift, R.A. and Decl. of the selected BCG, and the number of background source spectroscopic redshifts that were used to constrain $\cML$ and $\MSHM$ in this paper.

\startlongtable
\begin{deluxetable*}{lcccccc}
\tablecolumns{5}
\tablewidth{2\columnwidth}
\tablecaption{Strong Lensing Galaxy Clusters \label{table:sl_table}}
\tablehead{ \colhead{Galaxy Cluster} & \colhead{$\zl$} & \colhead{R.A.} & \colhead{Decl.} & \colhead{Detailed Lens Models} & \colhead{N($\zs$)} & Arcs/Model \\ [-8pt]
& & (J2000) & (J2000) & & & \colhead{Reference}}
\startdata
\\
SGAS \\
\hline
SDSS~J0108$+$0624 & 0.548 & 17.17511 & 6.41210 & L & 1 & a,b\\
SDSS~J0146$-$0929 & 0.447 & 26.73336 & $-$9.49792 & L & 2 & a,c\\
SDSS~J0150$+$2725 & 0.306 & 27.50355 & 27.42676 & L & 1 & a\\
SDSS~J0333$-$0651 & 0.573 & 53.26940 & $-$6.85635 & L & 1 & a\\
SDSS~J0851$+$3331 & 0.369 & 132.91194 & 33.51837 & L & 3 & a,d\\
SDSS~J0915$+$3826 & 0.396 & 138.91280 & 38.44952 & L & 2 & a,d,e\\
SDSS~J0928$+$2031 & 0.192 & 142.01889 & 20.52919 & L & 2 & a\\
SDSS~J0952$+$3434 & 0.357 & 148.16761 & 34.57947 & L & 1 & a,f\\
SDSS~J0957$+$0509 & 0.448 & 149.41330 & 5.15885 & L & 1 & a,d\\
SDSS~J1038$+$4849 & 0.431 & 159.68159 & 48.82159 & L & 3 & a,d\\
SDSS~J1050$+$0017 & 0.593 & 162.66637 & 0.28522 & L & 3 & a,g\\
SDSS~J1055$+$5547 & 0.466 & 163.76917 & 55.80647 & L & 2 & a,d\\
SDSS~J1110$+$6459 & 0.656 & 167.57386 & 64.99664 & L & 1 & a,c,h\\
SDSS~J1115$+$1645 & 0.537 & 168.76845 & 16.76058 & L & 2 & a,c,i\\
SDSS~J1138$+$2754 & 0.451 & 174.53731 & 27.90854 & L & 2 & a,d\\
SDSS~J1152$+$0930 & 0.517 & 178.19748 & 9.50409 & L & 1 & a\\
SDSS~J1152$+$3313 & 0.361 & 178.00077 & 33.22827 & L & 2 & a,d\\
SDSS~J1207$+$5254 & 0.275 & 181.89965 & 52.91644 & L & 1 & a,f\\
SDSS~J1209$+$2640 & 0.561 & 182.34877 & 26.67950 & L & 2 & a,d,j\\
SDSS~J1329$+$2243 & 0.443 & 202.39391 & 22.72106 & L & 1 & a,g\\
SDSS~J1336$-$0331 & 0.176 & 204.00035 & $-$3.52496 & L & 2 & a\\
SDSS~J1343$+$4155 & 0.418 & 205.88685 & 41.91763 & L & 1 & a,d,k\\
SDSS~J1420$+$3955 & 0.607 & 215.16680 & 39.91859 & L & 2 & a,d\\
SDSS~J1439$+$1208 & 0.427 & 219.79076 & 12.14043 & L & 2 & a\\
SDSS~J1456$+$5702 & 0.484 & 224.00368 & 57.03898 & L & 1 & a\\
SDSS~J1522$+$2535 & 0.602 & 230.71985 & 25.59097 & L & 1 & a\\
SDSS~J1531$+$3414 & 0.335 & 232.79429 & 34.24031 & L & 2 & a,d\\
SDSS~J1604$+$2244 & 0.286 & 241.04227 & 22.73858 & L & 1 & a\\
SDSS~J1621$+$0607 & 0.343 & 245.38494 & 6.12197 & L & 2 & a,d\\
SDSS~J1723$+$3411 & 0.442 & 260.90068 & 34.19948 & L & 2 & a,f\\
SDSS~J2111$-$0114 & 0.636 & 317.83062 & $-$1.23984 & L & 1 & a,d\\

\hline
\\
CLASH\\
\hline
Abell~383 & 0.189 & 42.01409 & $-$3.52938 & LTM.v2, NFW.v2 & 4 & l,n,s,t,u\\
Abell~611 & 0.288 & 120.23673 & 36.05656 & LTM.v2, NFW.v2 & 2 & l,v\\
MACS~J0329.7$-$0211 & 0.450 & 52.42321 & $-$2.19623 & L.v1, LTM.v2, NFW.v2 & 7 & l,m,n\\
MACS~J0416.1$-$2403 & 0.396 & 64.03808 & $-$24.06750 & See HFF & 37 & See HFF\\
MACS~J0429.6$-$0253 & 0.399 & 67.40003 & $-$2.88521 & L.v1, LTM.v2, NFW.v2 & 2 & l,m\\
MACS~J0717.5$+$3745 & 0.545 & 109.39855 & 37.75479 & See HFF & 8 & See HFF\\
MACS~J1115.9$+$0129 & 0.353 & 168.96628 & 1.49861 & L.v1, LTM.v2, NFW.v2 & 2 & l,m,n\\
MACS~J1149.5$+$2223 & 0.543 & 177.39875 & 22.39853 & See HFF & 7 & See HFF\\
MACS~J1206.2$-$0847 & 0.440 & 181.55064 & $-$8.80094 & LTM.v2, NFW.v2 & 4 & l,n,w,x\\
MACS~J1311.0$-$0310 & 0.494 & 197.75751 & $-$3.17770 & L.v1, LTM.v2, NFW.v2 & 1 & l,m\\
MACS~J1423.8$+$2404 & 0.545 & 215.94949 & 24.07846 & LTM.v2, NFW.v2 & 2 & l,y\\
MACS~J1931.8$-$2635 & 0.352 & 292.95684 & $-$26.57587 & L.v1, LTM.v2, NFW.v2 & 7 & l,m\\
MACS~J2129.7$-$0741 & 0.570 & 322.35879 & $-$7.69105 & L.v1, LTM.v2, NFW.v2 & 11 & l,m,o\\
MS~J2137$-$2353 & 0.315 & 325.06316 & $-$23.66114 & LTM.v2, NFW.v2 & 2 & l\\
RXC~J1347.5$-$1145 & 0.451 & 206.88261 & $-$11.75318 & L.v1, LTM.v2, NFW.v2 & 4 & l,m,n,p,q,r\\
RXC~J2129.7$+$0005 & 0.234 & 322.41649 & 0.08922 & L.v1, LTM.v2, NFW.v2 & 7 & l,m\\
RXC~J2248.7$-$4431 & 0.348 & 342.18321 & $-$44.53089 & See HFF (Abell~S1063) & 18 & See HFF (Abell~S1063)\\

\hline
\\
HFF\\
\hline
Abell~2744 & 0.308 & 3.58626 & $-$30.40017 & C.v4, C.v4.1, D.v4 & 26 & z,aa,ag,ah,ak,al,am\\
 & & & & D.v4.1, G.v4, K.v4 & & \\
 & & & & S.v4c, W.v4 & & \\
Abell~370 & 0.375 & 39.97133 & $-$1.58224 & B.v4, B.v4.1, C.v4 & 32 & z,ag,ai,aj,al,am\\
 & & & & D.v4, D.v4.1, G.v4 & & \\
 & & & & K.v4, S.v4, W.v4 & & \\
 & & & & W.v4.1 & & \\
Abell~S1063 & 0.348 & 342.18321 & $-$44.53089 & C.v4, C.v4.1, D.v4 & 18 & z,ab,af,ag,al,am\\
 & & & & D.v4.1, G.v4, K.v4 & & \\
 & & & & S.v4, W.v4, W.v4.1 & & \\
MACS~J0416.1$-$2403 & 0.396 & 64.03808 & $-$24.06750 & Cam.v4, C.v4, C.v4.1 & 37 & z,ae,ag,al,am\\
 & & & & D.v4, D.v4.1, G.v4 & & \\
 & & & & K.v4, S.v4c, W.v4 & & \\
MACS~J0717.5$+$3745 & 0.545 & 109.39855 & 37.75479 & C.v4, C.v4.1, D.v4 & 8 & z,ad,al,am\\
 & & & & D.v4.1, K.v4, S.v4c & & \\
 & & & & W.v4, W.v4.1 & & \\
MACS~J1149.5$+$2223 & 0.543 & 177.39875 & 22.39853 & C.v4, C.v4.1, D.v4 & 7 & z,ac,al,am\\
 & & & & D.v4.1, K.v4, S.v4c & & \\
 & & & & W.v4 & & \\

\hline
\\
RELICS\\
\hline
Abell~2537 & 0.297 & 347.09256 & $-$2.19212 & L.v1, G.v2 & 1 & an\\
Abell~2813 & 0.292 & 10.85271 & $-$20.62822 & L.v1 & 1 & au\\
Abell~3192 & 0.425 & 59.72531 & $-$29.92527 & L.v1 & 2 & ao\\
Abell~S295 & 0.300 & 41.35339 & $-$53.02932 & LTM.v2 & 1 & ap\\
CL~J0152.7$-$1357 & 0.833 & 28.18242 & $-$13.95515 & L.v1, LTM.v1, G.v2 & 1 & aq\\
MACS~J0025.4$-$1222 & 0.586 & 6.36415 & $-$12.37303 & LTM.v1 & 1 & ap\\
MACS~J0035.4$-$2015 & 0.352 & 8.85889 & $-$20.26229 & L.v1, G.v2 & 2 & au\\
MACS~J0257.1$-$2325 & 0.505 & 44.28647 & $-$23.43468 & L.v1, G.v2 & 1 & au\\
MACS~J0417.5$-$1154 & 0.443 & 64.39454 & $-$11.90885 & L.v2, G.v3 & 2 & ar\\
MACS~J0553.4$-$3342 & 0.430 & 88.33069 & $-$33.70754 & L.v1, G.v2 & 1 & au\\
MS~1008.1$-$1224 & 0.306 & 152.63455 & $-$12.66469 & L.v1 & 2 & au\\
PLCK~G004.5$-$19.5 & 0.540 & 289.27098 & $-$33.52236 & L.v1 & 5 & au\\
RXC~J0018.5$+$1626 & 0.546 & 4.63992 & 16.43787 & L.v1 & 2 & au\\
RXC~J0032.1$+$1808 & 0.396 & 8.03914 & 18.11561 & L.v1, LTM.v2, G.v2 & 1 & as\\
RXC~J0232.2$-$4420 & 0.284 & 38.06804 & $-$44.34669 & L.v1 & 1 & au\\
RXC~J2211.7$-$0350 & 0.397 & 332.94137 & $-$3.82895 & L.v1, G.v2 & 1 & an\\
SPT$-$CL~J0615$-$5746 & 0.972 & 93.96543 & $-$57.78011 & L.v1, LTM.v1 & 2 & at\\

\enddata
\tablenotetext{}{\textsc{\textbf{Strong Lensing Galaxy Clusters Included in this Work.}}  $\zl$ is the lens redshift of the galaxy cluster, R.A. and Decl. are the right ascension and declination of the selected BCG, respectively, and $N(\zs)$ is the number of multiply-imaged lensed background sources with spectroscopic redshifts that are used in this paper. The Detailed Lens Models lists indicate the name of the lens modeling teams or algorithms, and the versions which are utilized for the comparison in this work. A brief description of the samples can be found in \S\ref{sec:data}: SGAS (see \S\ref{subsec:sgas}), CLASH (see \S\ref{subsec:clash}), HFF (see \S\ref{subsec:hff}), and RELICS (see \S\ref{subsec:relics}).}
\tablecomments{\\[1pt]
Detailed Lens Models used in this work (see also \S\ref{sec:data}): \\
SGAS: L~$=$~\LENSTOOL\\
CLASH: L~$=$~\LENSTOOL; LTM~$=$~Light-Traces-Mass; NFW~$=$~LTM $+$ eNFW; .v1~$=$~version 1; .v2~$=$~version 2\\
HFF: B~$=$~Brada\v{c} (SWUnited); C~$=$~CATS (\Lenstool); Cam~$=$~Caminha (\Lenstool); D~$=$~Diego (WSLAP+); G~$=$~Glafic (GLAFIC); K~$=$~Keeton ({\tt{GRAVLENS}}); S~$=$~Sharon (\Lenstool); W~$=$~Williams (GRALE); .v4~$=$~version 4; .v4c~$=$~version 4 corrected; .v4.1~$=$~version 4.1\\
RELICS: G~$=$~GLAFIC; L~$=$~\LENSTOOL;  LTM~$=$~Light-Traces-Mass; .v1~$=$~version 1; .v2~$=$~version 2; .v3~$=$~version 3 \\
References: \\
a)~\citet{Sharon:20};
b)~\citet{Rigby:18};
c)~\citet{Stark:13};
d)~\citet{Bayliss:11a};
e)~\citet{Bayliss:10};
f)~\citet{Kubo:10};
g)~\citet{Bayliss:14};
h)~\citet{Johnson:17a}:
i)~\citet{Bayliss:12};
j)~\citet{Ofek:08};
k)~\citet{Diehl:09};
l)~\citet{Zitrin:15};
m)~\citet{Caminha:19};
n)~CLASH-VLT Rosati et al. (in prep);
o)~\citet{Huang:16};
p)~\citet{Ravindranath:02};
q)~\citet{Bradac:08a};
r)~\citet{Halkola:08};
s)~\citet{Smith:01};
t)~\citet{Newman:11};
u)~\citet{Richard:11};
v)~\citet{Newman:13};
w)~\citet{Ebeling:09};
x)~\citet{Zitrin:12a};
y)~\citet{Limousin:10};
z)~\citet{Johnson:14};
aa)~\citet{Zitrin:14};
ab)~\citet{Diego:16};
ac)~\citet{Jauzac:16};
ad)~\citet{Limousin:16};
ae)~\citet{Caminha:17};
af)~\citet{Karman:17};
ag)~\citet{Kawamata:18};
ah)~\citet{Mahler:18};
ai)~\citet{Strait:18};
aj)~\citet{Lagattuta:19};
ak)~\citet{Sebesta:19};
al)~\citet{Vega-Ferrero:19};
am)~\citet{Raney:20a};
an)~\citet{Cerny:18};
ao)~\citet{Hsu:13};
ap)~\citet{Cibirka:18};
aq)~\citet{Acebron:19};
ar)~\citet{Mahler:19};
as)~\citet{Acebron:20};
at)~\citet{Paterno-Mahler:18};
au)~RELICS public data release (see \S\ref{subsec:relics})
}
\end{deluxetable*}


\section{Lens Modeling And Einstein Radius} 
\label{sec:lens_modeling}

Strong lens modeling analyses use the positional and redshift measurements of lensed galaxies (arcs) as constraints to model the underlying mass distribution. There are a variety of well-established lensing algorithms that have been used extensively to study both the galaxy cluster and the magnified background universe. Below, we provide a brief description of the lensing algorithms that were employed to compute the publicly available detailed lens models used in our analysis. We also briefly describe the Einstein Radius mass estimate, and single-halo lens models, which were recently evaluated by \citet{Remolina:20} and \citet{Remolina:21}, respectively, as methods to quickly and effectively measure the mass at the core of strong lensing galaxy clusters. 

\subsection{Detailed Lens Models} 
\label{subsec:dlm}

Lensing algorithms are usually grouped into three categories: parametric, non-parametric, and hybrid, based on the parametrization of the modeled mass distribution. Parametric models utilize a combination of parametric functions to describe the mass distribution of the lens plane. Non-parametric or ``free-form'' algorithms make no assumption on the functional form of the mass distribution. Hybrid models are a combination of these two forms. The degree to which mass is assumed to be correlated with the observed light distribution also varies among the different algorithms.

The parametric models that are used in this work include: GLAFIC \citep{Oguri:10,Ishigaki:15,Kawamata:16}, {\tt{GRAVLENS}} \citep{Keeton:10,McCully:14}, and \LENSTOOL\ \citep{Kneib:96, Jullo:07, Jullo:09, Niemiec:20}. These algorithms use a variety of analytical mass distributions both for the cluster-scale dark matter halos and the contribution of the galaxy cluster members. Light-Traces-Mass (LTM; \citealt{Broadhurst:05, Zitrin:09, Zitrin:15}) assigns mass to a parameterized description of the light distribution, and LTM with elliptical NFW profiles (LTM$+$eNFW; \citealt{Zitrin:09,Zitrin:15}) combines this approach with analytical mass distributions as the parametric models. The ``free-form’' algorithms include Strong and Weak Lensing United (SWUnited; \citealt{Bradac:06,Bradac:09}) which performs an iterative minimization of a non-regular adaptive grid and GRALE \citep{Liesenborgs:06,Mohammed:14}, which uses a genetic algorithm to iteratively refine the mass distribution on a grid. Last, the hybrid algorithm Weak \& Strong Lensing Analysis Package (WSLAP+; \citealt{Diego:05, Diego:07, Diego:16}) is a non-parametric algorithm with the addition of a parametrized distribution for the cluster member contribution. Modeling algorithms also differ by their assumptions on the extent of correlation between light and mass. A variety of techniques are employed to explore the parameter space and determine the model that best reproduces the observed lensing configuration, and determine statistical uncertainties.

Detailed lens models (DLM) can be highly complex, adding the flexibility required for detailed studies of galaxy cluster properties, their surrounding environment, uncorrelated structure along the line-of-sight, the magnified background universe, and cosmology. This high complexity of the models relies on a large number of free parameters, requiring a large number of constraints, i.e., multiply-imaged lensed galaxies, whose availability becomes a limiting factor in the modeling process. The versatility of DLM also means the models are not unique and require care in the construction and evaluation; statistical assessments are employed to select between models (e,g., \citealt{Acebron:17, Paterno-Mahler:18, Lagattuta:19, Mahler:19}). High-fidelity lens models of galaxy clusters with rich strong lensing evidence require extensive follow-up observations, large investment of computational and human resources, and multiple iterations of the lensing analysis and modeling process to revise the models as new observational evidence becomes available (e.g., \citealt{Sharon:12, Johnson:14, Jauzac:15}).

To determine the statistical uncertainties of the public DLM used in this work, we use the ``range'' maps that are provided with them. The ``range'' maps are the same lensing products as the best-fit products, except they are derived from sets of parameters that sample the parameter space of each model, and provide a handle on how the variation in model parameters affects the lensing-derived projected mass density.  

\subsection{Single-Halo Lens Models} 
\label{subsec:shm}

The single-halo lens models (SHM) computed in this analysis follows \citet{Remolina:21}. We use \LENSTOOL\ to compute the SHM in one lens plane with a single cluster-scale dark matter halo. The mass distribution is parameterized using a dual pseudo-isothermal ellipsoid (dPIE, \citealt{Eliasdottir:07}) and no contribution from galaxy cluster members. Of the seven dPIE parameters ($\Delta\alpha$ and $\Delta\delta$ are the R.A. and Decl.; $\epsilon$ is the ellipticity; $\theta$ is the position angle; r$_{core}$ is the core radius; r$_{cut}$ is the truncation radius; and $\sigma$ is the effective velocity dispersion), only six are optimized as we set the truncating radius to a fixed $1500$ kpc as is typically done in DLM in the literature (note that this projected radius is also similar to the spashback radius; e.g., \citealt{Umetsu:17,Shin:19}). We use broad priors in the six free parameters of the dPIE potential: $-8\farcs0 < \Delta\alpha, \Delta\delta < 8\farcs0$ ; $0.0 < \epsilon < 0.9$ ;  $0^{\circ}  < \theta <180^{\circ}$ ; $50$ kpc $<$ r$_{core}$ $< 150$ kpc; and $500$ km/s $< \sigma < 1500$ km/s. The small number of free parameters calls for only a handful of constraints, with a minimum of 6 constraints required. This can be satisfied with as little as 4 multiple images of the same source, as each identified set of $n$ multiple images contributes $2n - 2$ constraints. With the image identification in hand (see \S\ref{subsec:arc_cat}), the models can be computed quickly and with limited human intervention. Generally, the SHM can be automatically computed once the cluster redshift, center initial position (e.g., the brightest cluster galaxy - BCG), and position and redshift of the arcs are measured.

\citet{Remolina:21} assess the scatter and bias associated with the mass estimated by this approach by comparing it to ``true'' mass from mock strong lensing images based on the Outer Rim \citep{Heitmann:19} cosmological simulation. They measure the single-halo aperture mass within the effective Einstein radius, $\MSHM$, from the projected mass distribution derived by the SHM, and compare it to  the mass from the simulated data, which they measure within the same radius. They find
an overall scatter of \RMSHMAllScatter\ with a bias of \RMSHMAllBias\ in $\MSHM$. When a quick visual inspection is performed and only the models that pass the inspection are used, the scatter and bias of $\MSHM$ improve to \RMPSHMAllScatter\ and \RMPSHMAllBias, respectively. The visual inspection is conducted in order to identify those single-halo lens models that fail to reproduce the observed lensing configuration and predict arcs in regions where no multiple images are found. 

The aperture within which the masses were measured in \citet{Remolina:21}, as well as in this work, is the effective Einstein radius (denoted as $e\theta_{\rm E}$ in \citealt{Remolina:21}), defined as the radius of a circle with the same area enclosed by the tangential critical curve of the SHM. The critical curves are derived from the convergence and shear outputs of the best-fit SHM. We use a notation of $\RSHM$ instead of $e\theta_{\rm E}$ in order to reduce confusion with other notations used in this paper.

\subsection{Einstein Radius} 
\label{subsec:er}

The  mass enclosed by the Einstein radius, $\ML$, is a quick method to estimate cluster core mass, where strong lensing is detected:

\begin{equation}
\label{eq:m_er}
	M(<\theta_E)=\Sigma_{cr} (\zl,\zs) \  \pi \ [\dl(\zl)\theta_E]^2,
\end{equation}

\noindent where $\Sigma_{cr} (\zl,\zs)$ is the critical surface density, $\dl(\zl)$ is the angular diameter distance from the observer to the lens, $\zl$ is the lens redshift, $\zs$ is the background source redshift, and $\ER$ is the Einstein radius. The main assumption of this method is that the projected mass distribution of the lens is circularly symmetric \citep{Narayan:96, Schneider:06a, Kochanek:06, Bartelmann:10, Kneib:11}. In this method, a crude estimate of the Einstein radius is obtained from the occurrence of arcs around the center of the lens, e.g., by minimizing the quadrature sum of the difference between the arc positions and the nearest point to them on the circle.

\citet{Remolina:20} quantified the scatter and bias of the mass enclosed by the Einstein radius method using mock lensed images from the Outer Rim \citep{Heitmann:19} simulations. They find that the scatter and bias increase with deviation from spherical symmetry and with the estimated $\theta_E$, and introduced empirical corrections to de-bias the results and reduce the scatter.

The empirical correction was calibrated for estimated Einstein radii of $\theta_E \leq 30\farcs0$, and for different centering assumptions. The corrected mass enclosed by the estimated Einstein radius, $M_{corr}(<\theta_{\rm E})$, is reported to have no bias, and a scatter of \RcMLAllScatter\ for the quadratic (\RcMLAllLinScatter\ for the linear) corrected masses. Only the identified tangential arcs are used in this method.

We use the same methods as \citet{Remolina:20} for estimating the Einstein radius and calculating the enclosed mass using \autoref{eq:m_er}. However, in the rest of the paper we denote the estimated Einstein radius used in this method as $\eER$ instead of $\theta_E$, to highlight its deviation from the ``true'' or ``effective'' Einstein radii of the lens, and reduce confusion with other notations used in this paper. The empirically-corrected mass estimated by this method is denoted $\cML$ hereafter.

Following the recommendation and procedures established by \citet{Remolina:20}, when applying this method to the observational data we use the BCG of the galaxy cluster as our fixed center.

The projected arc radii in this work extend beyond the calibrated range (see \textit{left} panel of \autoref{fig:ER_eER_dist}). We therefore use caution when applying this method to our sample, and investigate different choices in its application at large estimated Einstein radii. We apply the quadratic empirical correction for $\eER \leq 30\farcs0$, as recommended by \citet{Remolina:20}, and the linear empirical correction for the rest. In addition to the full sample, we report results for a subsample of $\eER \leq 20\farcs0$, which is better represented by the simulated data used by \citet{Remolina:20} to calibrate the method.


\section{Methodology} 
\label{sec:methodology}

In the following section, we describe the input constraints needed to compute $\cML$ and $\MSHM$. Following the work by \citet{Remolina:20} and \citet{Remolina:21}, we compute the core mass for the sample of strong lensing galaxy clusters analysed in this work.

\subsection{Arc Catalogs}
\label{subsec:arc_cat}

We use the lensing constraints (arcs) that were identified and listed with the public lens models. For this work, we only use constraints with spectroscopic redshifts (reference for the arc catalogs are given in \S\ref{sec:data}). We inspect the lensed galaxies and determine if they are tangential or radial arcs depending on the direction of their distortion. Only the tangential arcs are used in the fits for the $\ML$ method, but all of the arcs are included when computing the single-halo lens models.

\subsection{BCG Selection}
\label{subsec:bcg}

The position of the BCG serves as the initial position for the cluster-scale dark matter halo in the single-halo lens models and as the fixed center in the Einstein radius mass estimate. The BCGs were selected by their magnitude from a cluster member catalog (see \citealt{Postman:12} and \citealt{Fox:21}) and then confirmed by visual inspection.

\subsection{Computing $\MSHM$ and $\cML$}
\label{subsec:ML_MSHM}

\paragraph{SHM method} Using the catalog of the arcs and the selected BCG, we compute the single-halo lens models, compute $\RSHM$ from the SHM critical curves, and measure the aperture mass within a radius of $\RSHM$. As noted in \S\ref{subsec:shm}, $\RSHM$ is defined as the radius of a circle with the same area as the tangential critical curve. We compute a SHM for each set (also known as ``family'') of multiply-imaged background sources. The resulting SHM outputs (projected mass density, convergence, and shear) are used to compute $\MSHM$ as described in \S\ref{subsec:shm}.

There are cases where a galaxy cluster has multiple arc families although none of the individual families satisfy the minimum number of $6$ constraints needed (the total number of constraints for a given model is $\Sigma(2n_i-2)$, where $n_i$ is the number of constrains for background source $i$). We therefore compute one SHM for each cluster that uses all the families as constraints, thus the minimum number of constrains needed is attained. For these models, the SHM outputs are computed for a source redshift of $\zs = 2.0$. All the SHM are inspected and only the ones that pass the quick visual inspection are used in our analysis. From the total of $67$ clusters, $62$ ($29$ SGAS, $15$ RELICS, $6$ HFF, and $12$ CLASH) have enough constraints to compute a SHM, i.e., $6$ or more constraints. Following the visual inspection, only $54$ ($23$ SGAS, $13$ RELICS, $6$ HFF, and $12$ CLASH) clusters remain in our analysis.

\begin{figure*}
    \center
    \includegraphics[width=1\textwidth]{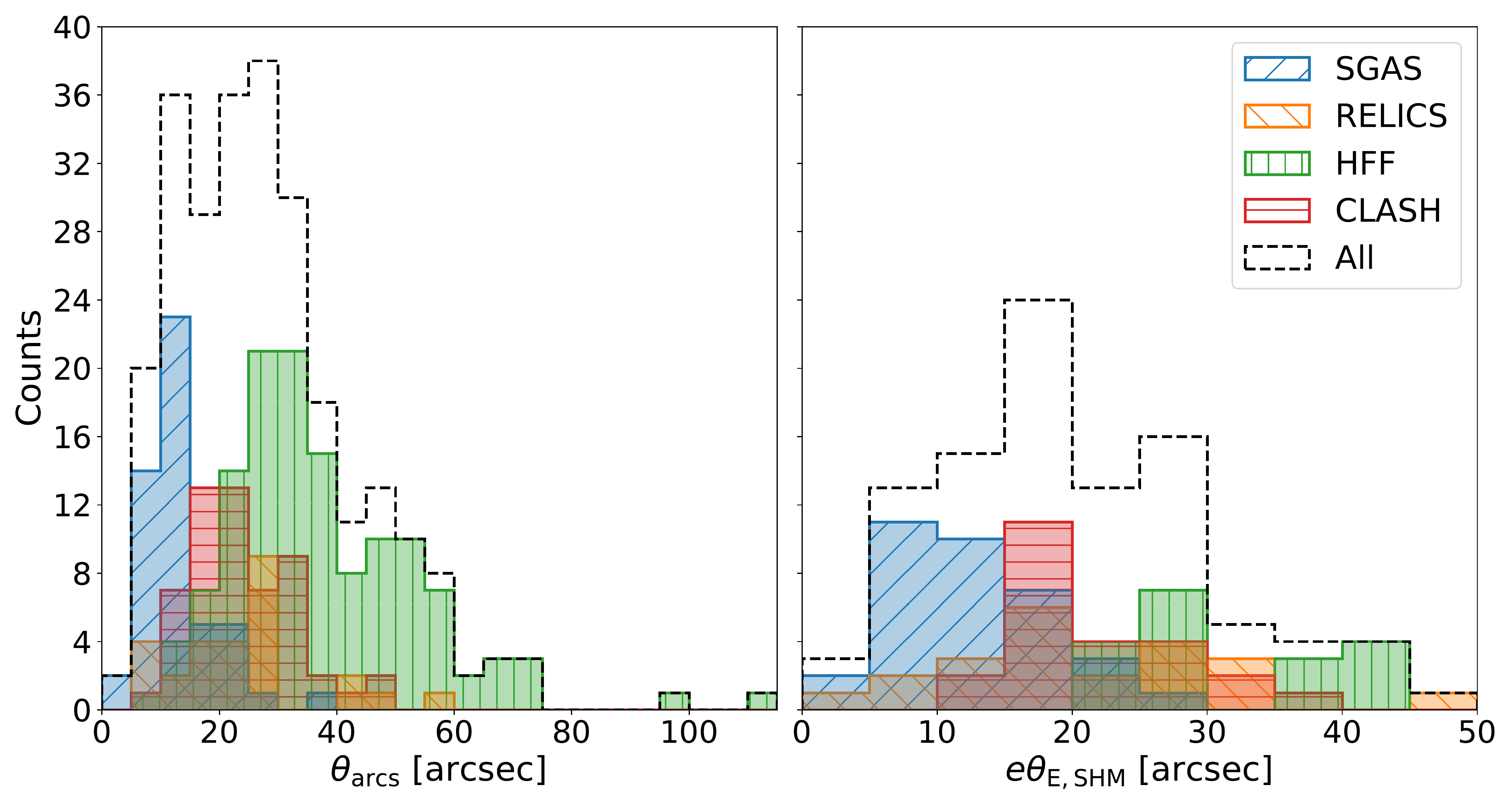}
    \caption{\textsc{\textbf{Distribution of $\eER$ and $\RSHM$ as measured from the two different mass estimate methods.}} The approximate Einstein radius ($\eER$; \textit{left} panel) is measured from the geometric fit of a circle that minimizes the quadrature sum of distances between the tangential arc positions and the nearest points to them on the circle of a single background source.  The SHM-derived effective Einstein radius ($\RSHM$; \textit{right} panel) is measured as the radius of a circle with the same area enclosed by the tangential critical curve of the single-halo lens models. This figure shows only results from SHMs that passed the visual inspection. Both $\eER$ and $\RSHM$ have units of arcseconds. The black dashed line represents the total counts and the colors denote the counts from the four different surveys of strong lensing galaxy clusters. As expected from the selection functions of these samples, SGAS clusters have lower $\eER$ and $\RSHM$, followed by the RELICS, CLASH, and HFF galaxy clusters. The deep observation and extensive followup of the six HFF clusters result in a large number of lensed sources with spectroscopic redshifts, extending to large cluster-centric radii, which is reflected in the distribution of $\eER$. SGAS, CLASH (except for those that are also part of HFF), and RELICS have only a few lensed sources per cluster with spectroscopic redshifts, and are found at smaller cluster-centric distances.}
    \label{fig:ER_eER_dist}
\end{figure*}

We plot the distribution of effective radii, $\RSHM$, measured from the single-halo lens models that pass the visual inspection in the \textit{right} panel of \autoref{fig:ER_eER_dist}. The distribution of $\RSHM$ generally follows the number of clusters in each survey, as most clusters only have one or two independent SHM that could be computed and pass the visual inspection. We note that while the depth of the HFF data leads to an unprecedented number of strongly lensed galaxies overall, many of the arc families do not have four or more secure multiple images each. We find that SGAS models occupy the lower end of the $\RSHM$ distribution, followed by RELICS, CLASH, and HFF. The distributions of effective Einstein radii measured from the SHM emphasizes the difference in the selection function of the strong lensing sample, as CLASH, HFF, and RELICS attempted to select clusters with large lensing cross section, to increase the chances of observing magnified high redshift galaxies.
 
\paragraph{Einstein radius method} Utilizing the same catalog of arcs and BCG positions, we geometrically fit each arc family with a circle that minimizes the quadrature sum of distances between the tangential arc positions and the nearest points to them on the circle, following \cite{Remolina:20}. The resulting radius, $\eER$, is assumed to be an approximation of $\theta_E$ in \autoref{eq:m_er}. We measure at least one $\eER$ per galaxy cluster. The measured $\eER$ is then used to compute $\cML$ as described in \S\ref{subsec:er}, using \autoref{eq:m_er}, and the empirical correction from \citet{Remolina:20}. 

We plot the distribution of all $\eER$ in the \textit{left} panel of \autoref{fig:ER_eER_dist}. Unlike the SHM case, the $\cML$ can be computed for any number of multiple images of a given lensed source, resulting in a $\cML$ measurement for each strongly-lensed source with spectroscopic redshift. The deep observations and extensive spectroscopic followup of the six HFF clusters resulted in a large number of lensed sources with spectroscopic redshifts, which extend to large cluster-centric radii. In all the other fields, where only a few lensed sources per cluster have spectroscopic redshifts, the number of measurements is driven by the number of clusters in each sample, and the identified sources have smaller cluster-centric distances. 

\subsection{Statistics}
\label{subsec:stats}

Depending on the number of arcs and arc families available for each method, each cluster enables up to $37$ $\eER$ measurements and up to six SHM. The measurements in each cluster are expected to be correlated, and their distribution can inform the statistical uncertainty. On the other hand, individual clusters are independent of each other. 

We follow \citet{Remolina:20,Remolina:21} and build a statistical sample for each method ($\ML$ and SHM) to take into account multiple mass estimates for a single galaxy cluster and set the statistical weight for each cluster equal to one. 

Depending on the number of available arc families, a given galaxy cluster may have more than one SHM or Einstein radius mass estimates. 
For the SHM mass estimate, we select at random one $\MSHM$ from the available single-halo lens models for each cluster. This process is repeated $1,000$ times per cluster, leading to a sample of {$62,000$} points from all SHM, of which {$54,000$} $\MSHM$ points are ones that passed the quick visual inspection.

A similar process is employed for the Einstein radius mass estimate. For each cluster we select at random one of its available arc families, 
and select a $\eER$ by sampling from a normal distribution centered on the fiducial $\eER$ measurement and a standard deviation equal to the uncertainty from the radius fit. We then calculate the relevant mass from \autoref{eq:m_er}. Again, we repeat the process $1,000$ times per cluster leading to a sample of {$67,000$} $\cML$ points.

For comparison of each of these sample points to $\MDLM$, the uncertainty in the DLM mass is accounted for by drawing from a normal distribution centered on the best-fit DLM and with standard deviation computed from the DLM ``range'' maps. If a cluster has more than one detailed lens model (see \S\ref{sec:data}), one DLM was selected at random for each of the 1000 sampling points.


\section{Analysis of Results} 
\label{sec:results}

In the following section, we compare the galaxy cluster core mass measurements obtained by the quick methods, $\cML$ and $\MSHM$, to the mass enclosed by the respective mass apertures from the detailed lens models, $\MDLM$. We evaluate the results against several properties of the lens system, and compare the scatter to that expected from simulations \citep{Remolina:20, Remolina:21} and from the statistical uncertainty of detailed lens models. In this work the scatter is defined as half of the difference between the $84$th and $16$th percentiles. The bias is determined from the median of the distribution.

\subsection{Mass Enclosed by the estimated Einstein Radius, $\cML$}
\label{subsec:ML}

\begin{figure*}
\center
\includegraphics[width=0.49\textwidth]{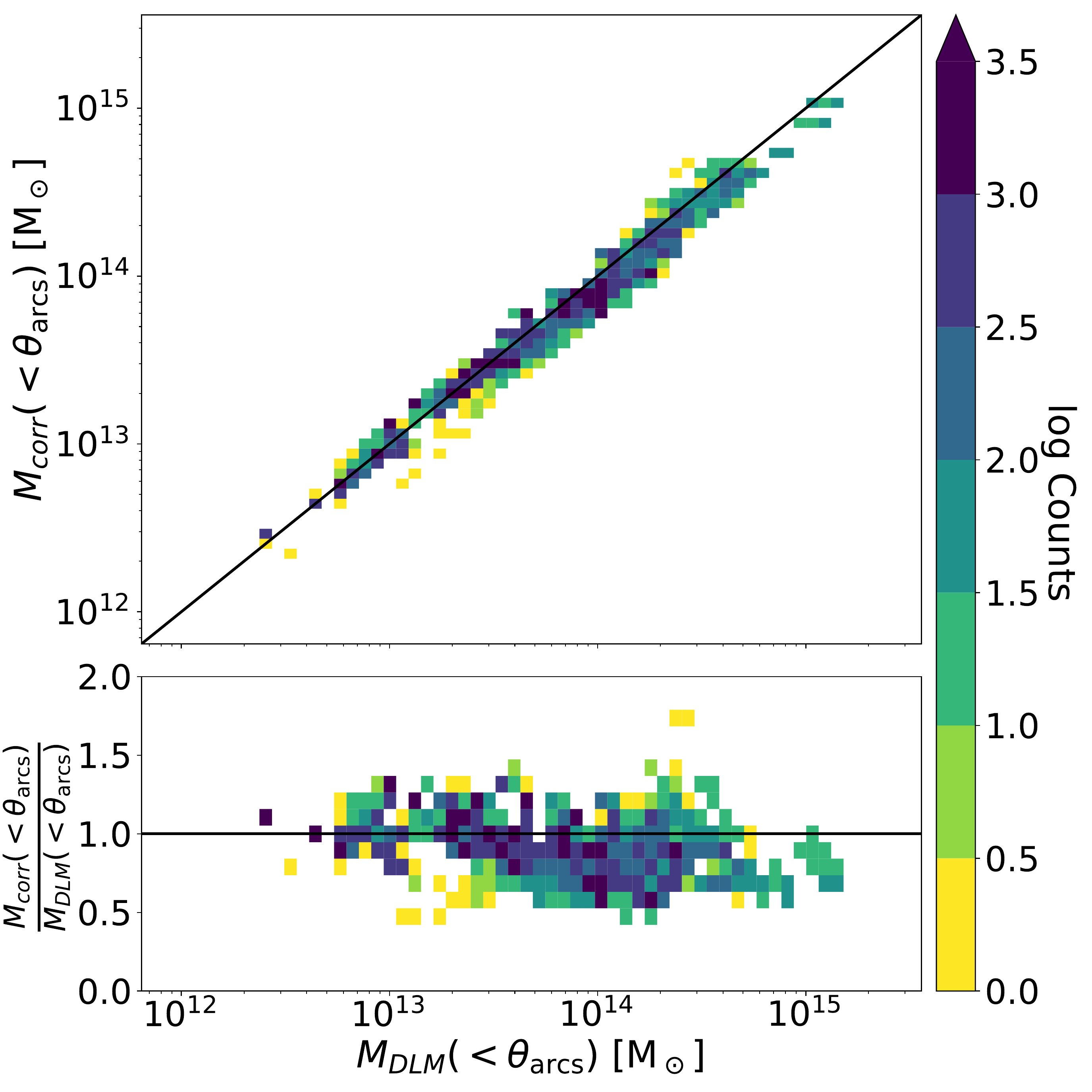}
\includegraphics[width=0.49\textwidth]{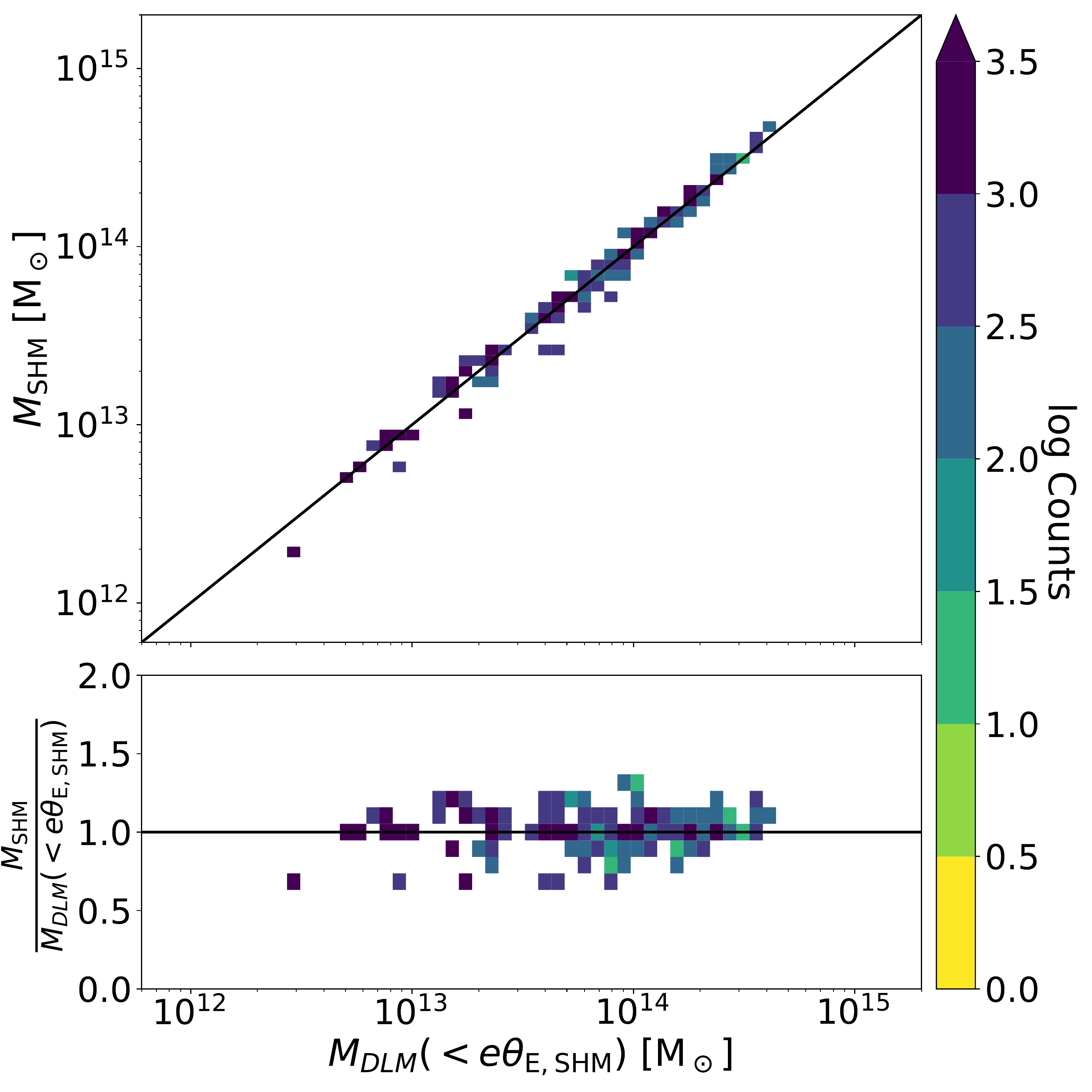}
\caption{\textsc{\textbf{Mass Comparison Between the Efficient Mass Estimates, $\cML$ and $\MSHM$, and the DLM, $\MDLM$.}} The \textit{left} panels are for the mass enclosed by the approximate Einstein radius, $\cML$, and the \textit{right} panels are for the SHM that passed the visual inspection, $\MSHM$. The \textit{top} plot shows the direct comparison between the masses and the \textit{bottom} plot is the ratio of the mass measurements. The total number of counts are the $62,000$ and $52,000$ sampled data points for the $\cML$ and $\MSHM$, respectively (see \S\ref{subsec:stats}). The black lines indicate the one-to-one line, where $\cML$ or $\MSHM$ equal $\MDLM$. We find that the distribution of $\cML$ is biased low particularly at large $\MDLM$ and the distribution of $\MSHM$ is slightly biased high.}
\label{fig:mass_comparison}
\end{figure*}

In \textit{left} panels of \autoref{fig:mass_comparison}, we plot the direct comparison between the corrected mass from the Einstein radius method, $\cML$, and the mass enclosed by the same aperture from the best-fit detailed lens model, $\MDLM$, for all clusters. We measure an overall scatter of \cMLAllScatter\ and bias of \cMLAllBias\ in $\cML$ compared to $\MDLM$. We find that the distribution is biased low, particularly at large $\MDLM$ values. The observed negative bias is reduced when excluding systems with large estimated Einstein radius ($\eER > 20\farcs0$). For the subsample of $\eER \leq 20\farcs0$ the scatter is \cMLSmallERScatter\ and the bias is \cMLSmallERBias.  This bias could be possibly addressed by extending the work of \citet{Remolina:20} to larger radii, by using simulations that include lower magnification lensed sources at larger cluster-centric distances.

\subsection{Mass Estimate from Single Halo Lens Models, $\MSHM$}
\label{subsec:MSHM}
We assess the results of the entire SHM sample, and the results of the subsample of models that passed the visual inspection. In the \textit{right} panels of \autoref{fig:mass_comparison}, we plot the direct comparison between the aperture mass of the SHM that passed the visual inspection, $\MSHM$, and the mass enclosed by the same aperture in the best-fit DLM, $\MDLM$. For the entire SHM sample, we measure an overall scatter of \MSHMAllScatter\ and a bias of \MSHMAllBias. For the SHM that passed the quick visual inspection, we measure an overall scatter of \MPSHMAllScatter\ and a bias of \MPSHMAllBias\ between $\MSHM$ and $\MDLM$. Similar to \citet{Remolina:21}, we find that the visual inspection helps decrease the scatter and bias between $\MSHM$ and $\MDLM$.

\subsection{Analysis of Systematics}
\label{subsec:systematics}

\begin{figure*}
\center
\includegraphics[width=0.95\textwidth]{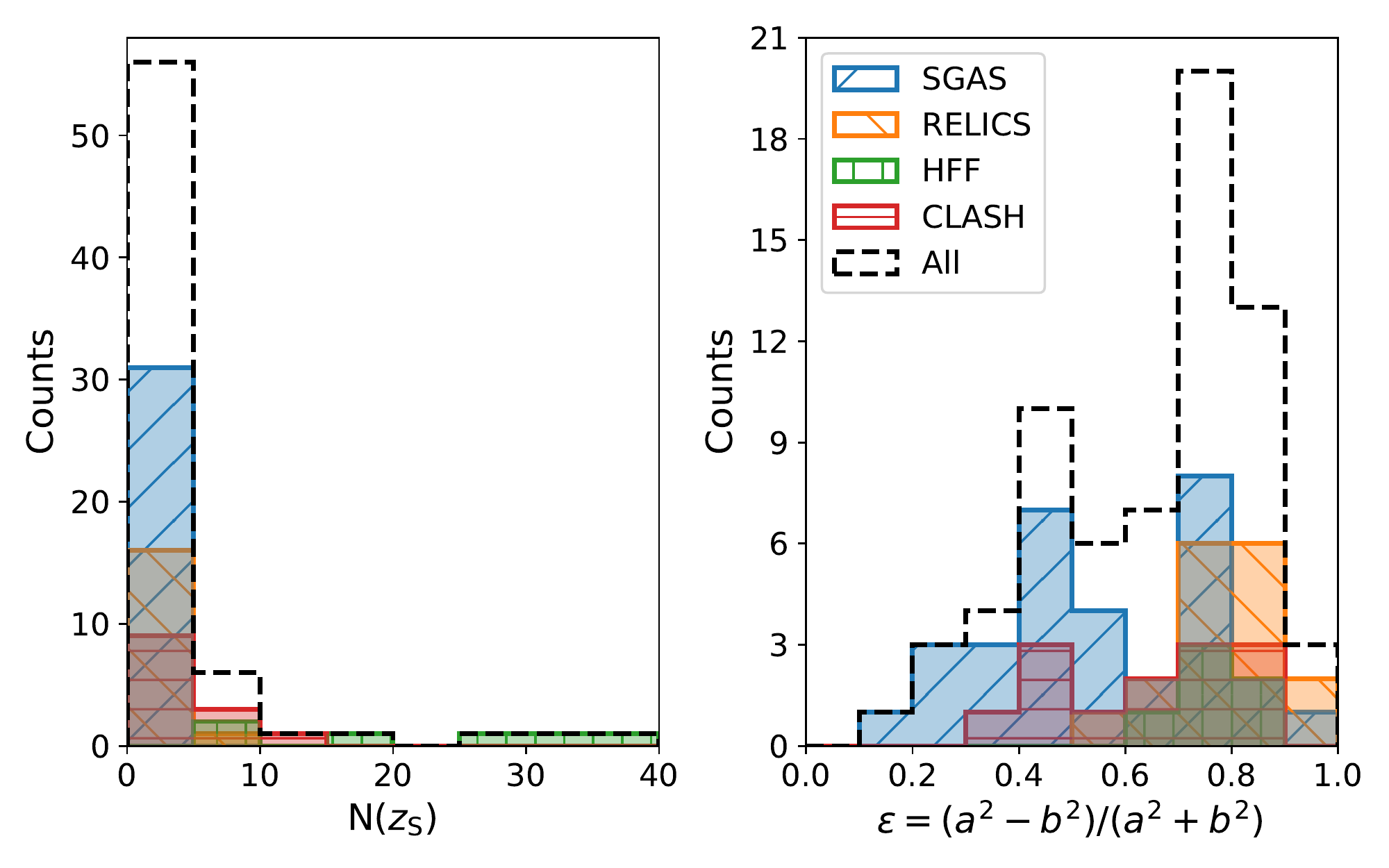}
\caption{\textsc{\textbf{Distribution of the Total Number of Multiply-imaged Lensed Sources with Spectroscopic Redshifts and Galaxy Cluster Deviation from Circular Symmetry.}} The total number of multiply-imaged background sources with spectroscopic redshifts per galaxy cluster, N($\zs$), is shown in the \textit{left} panel (see also \autoref{table:sl_table}). The distribution is indicative of the extensive observational and spectroscopic investment by the community in rich lensing clusters like the HFF. The deviation from circular symmetry is encoded in the ellipticity of the DLM critical curve, $\epsilon = (a^2-b^2)/(a^2+b^2)$, where $a$ and $b$ are the semi-major and semi-minor axes of an ellipse fit to the tangential critical curve, for a source redshift $\zs = 2.0$, computed from best-fit detailed lens models. When multiple DLMs exist for a galaxy cluster, the median $\epsilon$ is used. We find that the distribution matches expectations with the HFF having well reported complex and elongated mass distributions. We find that all the galaxy clusters from HFF and RELICS have $\epsilon > 0.5$, while CLASH and SGAS clusters are the only samples with some galaxy clusters with $\epsilon < 0.5$.}
\label{fig:nspec_ell_dist}
\end{figure*}

In this subsection, we discuss possible correlations between the scatter in the efficient mass estimates with the aperture radii within which they are measured ($\eER$ or $\RSHM$), the total number of multiply-imaged lensed background sources with spectroscopic redshifts available for each lens (N($\zs$)), and the galaxy clusters deviation from circular symmetry ($\epsilon$). The distributions of these properties in our sample are shown in \autoref{fig:ER_eER_dist} and \autoref{fig:nspec_ell_dist}, and briefly discussed below.

As can be seen in \autoref{table:sl_table} and the \textit{left} panel of \autoref{fig:nspec_ell_dist}, most of the clusters in our sample have five or fewer multiply-imaged background sources with spectroscopic redshifts, with the HFF and CLASH samples dominating the high-N($\zs$) end. The distribution of N($\zs$) is indicative of, and stems from, the extensive observational and spectroscopic efforts by the community in these fields. Clusters with $N(\zs)>5$ are shown to have highly accurate DLM \citep{Johnson:16}, and enable DLM with sufficient flexibility to describe complex mass distributions. We therefore compare clusters that fall within three broad bins: $N(\zs)=1$; $2 \leq N(\zs) \leq 5$; and $N(\zs)\geq 6$.

The deviation from circular symmetry of each cluster lens is estimated from their best-fit DLM. We compute the tangential critical curve for a background source redshift of $\zs = 2.0$ and fit an ellipse using the technique described in \citet{Fitzgibbon:96}. The resultant ellipticity adopts the following form: $\epsilon = (a^2-b^2)/(a^2+b^2)$, where $a$ and $b$ are the semi-major and semi-minor axes of the fitted ellipse, respectively. If multiple DLM are available for a particular galaxy cluster, the median $\epsilon$ is used. The distribution of $\epsilon$ (\textit{right} panel of \autoref{fig:nspec_ell_dist}) matches our expectation, with the complex and elongated structures of the HFF and RELICS galaxy clusters resulting in high values of $\epsilon$. We find $\epsilon < 0.5$ values only in CLASH and SGAS clusters.

\begin{figure*}
\center
\includegraphics[width=0.95\textwidth]{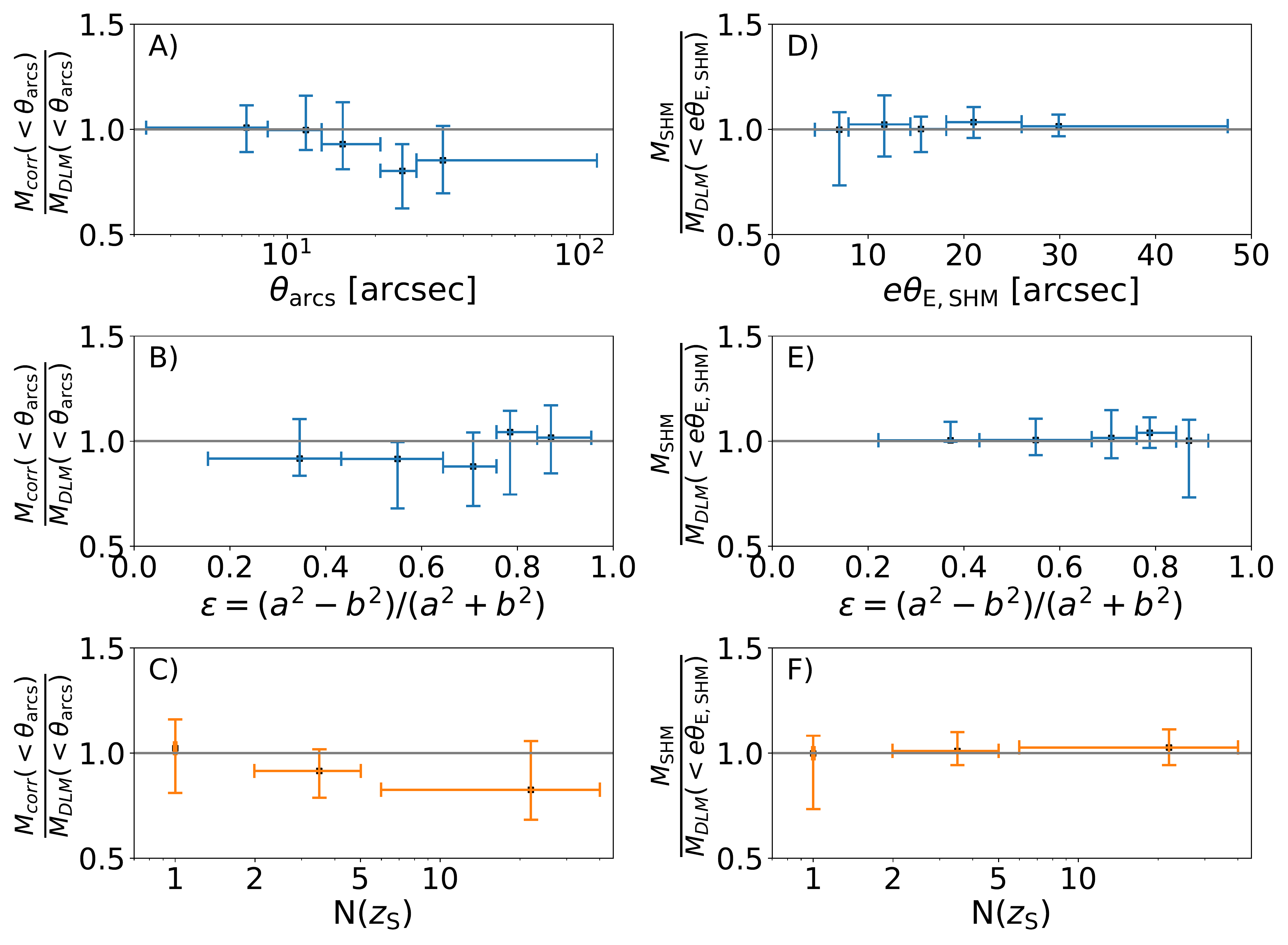}
\caption{\textsc{\textbf{Mass Ratio, $\cML / \MDLM(<\eER)$ and $\MSHM / \MDLM(<\RSHM)$, Binned by Radii, Deviation from Circular Symmetry, and Number of Background Source with Spectroscopic Redshift.}} The mass ratio between the efficient mass estimates and the mass from the detailed lens models, $\cML / \MDLM(<\eER)$ (\textit{left} panels) and $\MSHM / \MDLM(<\RSHM)$ (\textit{right} panels) binned by the approximate Einstein radius, ($\eER$, panel A), the SHM aperture ($\RSHM$, panel D), the total number of multiply-imaged lensed background sources with spectroscopic redshifts (N($\zs$), panels B and E), and the deviation from circular symmetry ($\epsilon$, panels C and F). The bins for $\eER$, $\RSHM$, and $\epsilon$ each have an equal number of points from the statistical samples of $67,000$ and $54,000$, see \S\ref{subsec:stats}. The N($\zs$) is divided into three bins: N($\zs$) $= 1$; $2 \leq$ N($\zs$) $\leq 5$; and N($\zs$) $\geq 6$, and do not have the same number of points per bin. The symbols indicate the median of the distribution and the error bars in the horizontal and vertical direction indicate the bin range and the scatter (the $16$th and $84$th percentile), respectively. We observe an overall negative bias in $\cML$ in panels A, B, and C. We find a zero bias in the first two $\eER$ bins, where this method is well-calibrated, and a negative bias at higher radii (panel A). The larger bias in high $N(\zs)$ bins reflects the difficulty of this single-component mass estimate to reconstruct the DLM complexity that is enabled by a large number of lensing constraints. In panels D, E, and F, we find that the mass ratio has little bias across all systematics we explore. We find a slight trend in the scatter in panels D and E, where the scatter decreases with increasing $\RSHM$ (panel D) and decreasing $\epsilon$ (panel E).}
\label{fig:bin_systematics}
\end{figure*}

\autoref{fig:bin_systematics} shows the mass ratio between the efficient mass estimates and the DLM. Results from the mass enclosed by the Einstein radius method are shown in the \textit{left} panels, and SHM in the \textit{right} panels. In the \textit{top} panels, we plot the mass ratios against the respective aperture within which they are measured, $\eER$ (\textit{left}) and $\RSHM$ (\textit{right}). The galaxy cluster deviation from circular symmetry, $\epsilon$, is shown in the \textit{middle} panels, and the number of multiply-imaged lensed background sources with spectroscopic redshifts, N($\zs$), in the \textit{bottom} panels. In the radii and the ellipticity panels, we use five bins with equal number of statistical sample points, $67,000$ and $54,000$ for $\cML$ and $\MSHM$, respectively. The N($\zs$) sample is divided into three non-uniform bins as described above. The symbols indicate the median of the distribution and the error bars in the horizontal and vertical direction indicate the bin range and the scatter (the $16$th and $84$th percentile), respectively.

In panel A of \autoref{fig:bin_systematics}, we find an indication of a decreasing trend in the mass ratio with increasing $\eER$. The last bin of $\eER$ is just consistent with a mass ratio of $1.0$, which we attribute to the change of the empirical correction from quadratic to linear for $\eER > 30\farcs0$ (see \S~\ref{subsec:er}). We confirm that in the region where the empirical correction was calibrated, $\eER<20\farcs0$, there is no bias. The negative bias observed at large $\eER$ could possibly be addressed by extending the work of \citet{Remolina:20} to larger cluster-centric radii and lower magnification. In panel B, we identify that while all ellipticity bins are consistent with a mass ratio of $1.0$, the lower $\epsilon$ bins have a small negative bias, while the opposite is observed for large $\epsilon$. The large number of galaxy clusters with large ellipticities that include many arcs with small $\eER$ may explain the trends seen in panels A and B. Last in panel C, we find that while all the three bins are consistent with a mass ratio of $1.0$, a trend of larger negative bias in the bins with higher N($\zs$) is identified. The highest bin highlights the Frontier Fields clusters and three CLASH clusters. The large number of constraints allows for highly flexible and complex detailed lens models. These galaxy clusters are also well known to have complex mass distributions which are not well represented by a circularly symmetric mass distribution. In addition, the extensive deep observations and follow-up work, has allowed identification of strong lensing evidence at large cluster-centric distance explaining the negative bias.

We find that $\MSHM$ (panels D, E, and F) has a very low bias and no trend is identified between the bias and $\RSHM$, $\epsilon$, and N($\zs$). In plot D, we observe a trend of decreasing scatter with increasing aperture radius. In panel E, we also identify a trend of increasing scatter  with increasing $\epsilon$. This trend matches our expectations as highly elongated and complex mass distributions represented by an elongated critical curve will have a larger uncertainty when applying the single-halo lens models, as these models are less complex than detailed lens models.

\subsection{Comparison to the Statistical Uncertainty of the Detailed Lens Models}
\label{subsec:MDLM}

\begin{figure}
\includegraphics[width=0.5\textwidth]{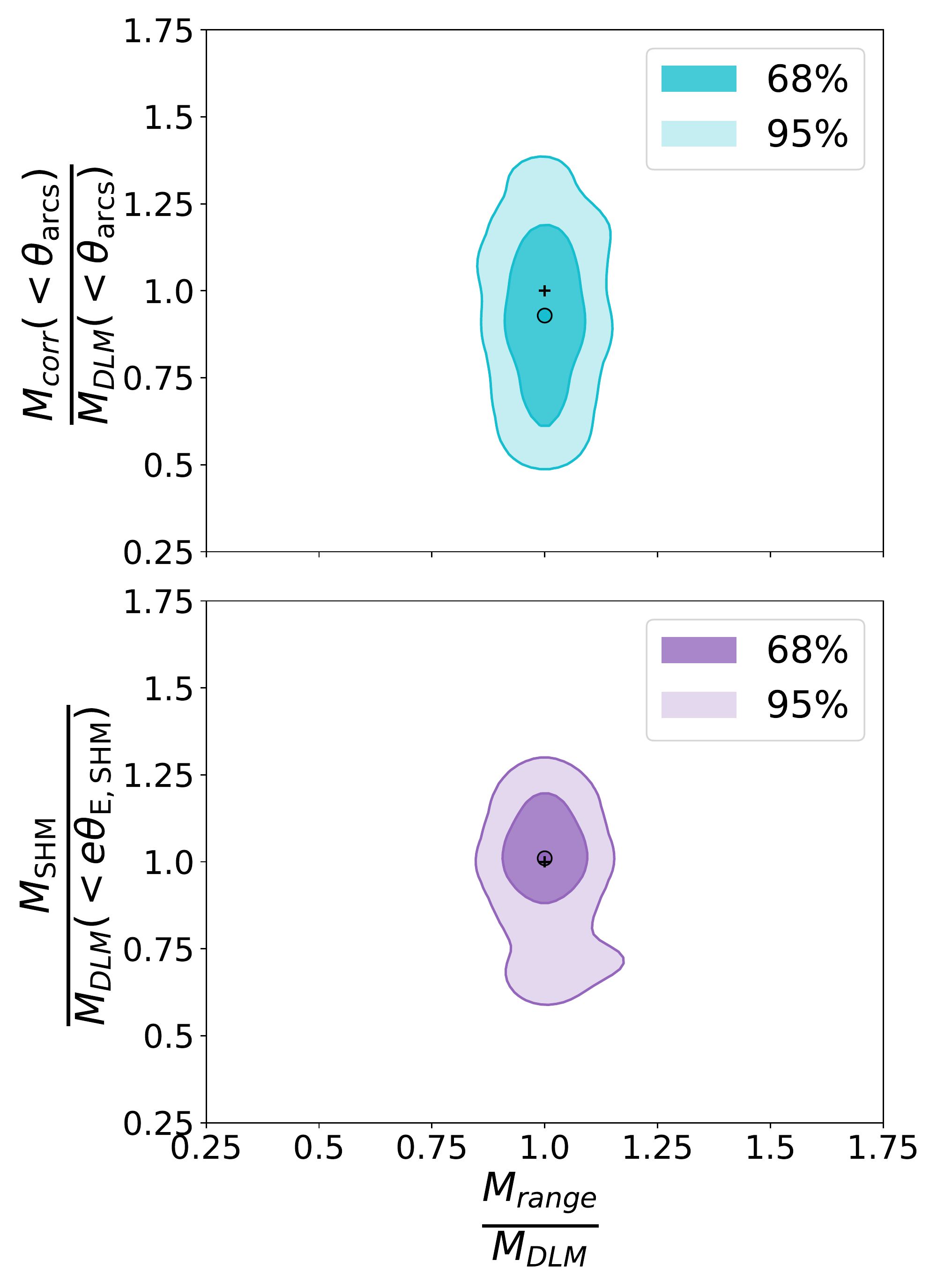}
\caption{\textsc{\textbf{The Scatter of the Efficient Mass Estimate Methods, Compared to the Statistical Uncertainty of the Detailed Lens Models.}} We plot the mass ratio between the mass estimate and the best-fit detailed lens model against  the statistical scatter of the detailed lens models, $M_{range}/M_{DLM}$, derived from the ratio of publicly available ``range'' maps and best-fit detailed lens model. The crosses stand for the point (1.0,1.0) and the open circles indicate the median of the distributions. Results for $\cML$  are shown in the \textit{top} panel, and for $\MSHM$  in the \textit{bottom} panel, smoothed by a kernel of $5\%$. The black cross indicates the location where $\cML$ and $\MSHM$ equal $\MDLM$.}
\label{fig:dlm_scatter_comp}
\end{figure}

To contextualize the scatter of the mass estimates assessed in this paper, we review it against the uncertainty typically attributed to detailed lens models. We plot in \autoref{fig:dlm_scatter_comp} the overall scatter in the $\cML$  and $\MSHM$ measurements against the statistical uncertainty of the detailed lens models, $M_{range}/M_{DLM}$, derived from the ratio of the ``range'' maps and the best-fit DLM. The statistical uncertainty of the detailed lens models is computed in the same way as the scatter (see \S\ref{sec:results}), except the uncertainty of each data point is drawn from the publicly available ``range'' maps provided by the lensing teams, and represents a statistical sampling of the parameter space, typically using MCMC. The aggregated statistical uncertainty over the entire sample from the detailed lens models is  $\sigma(\MDLM)=$ \DLMstatsig.

However, the statistical DLM modeling uncertainty is likely underestimated. Comparing models of two simulated clusters that were computed by different DLM algorithms, \citet{Meneghetti:17} conclude that detailed lens models are reliable when recovering the enclosed mass in the inner $100\farcs0$ with a scatter of less than $10\%$. In a recent comparison between DLM algorithms, \citet{Raney:20b} show that while the mass measured by the detailed lens models is reliable, the statistical uncertainty reported by the lensing algorithms underestimates the systematic uncertainty. \citet{Raney:20b} estimate the systematic uncertainty at $\sim 5\%$ for a circularly-averaged mass computed from the most recent versions (v4) of the HFF lens models.

\subsection{Comparison between Observations and Simulations}
\label{subsec:obs_vs_sim}

\citet{Remolina:20, Remolina:21} measured the scatter and bias of $\cML$  and $\MSHM$ against the ``true'' mass from simulations. To compare the scatter found in this work to \citet{Remolina:20, Remolina:21}, we need to account for the fact that detailed lens models are an observable measurement and while reliable are not the absolute truth. The expected scatter should therefore be a combination of the intrinsic scatter of the mass estimate, as measured from simulations, and the scatter attributed to the DLM measurement.

We note that the scatter between the three mass estimates ($\cML$, $\MSHM$, and $\MDLM$) may be correlated. To fully characterize the correlations between the masses will require the computation of detailed lens models for a large sample of simulated strong lensing galaxy clusters, which awaits new large cosmological simulations with baryonic information and will require an extensive amount of computational and human resources. 

With this in mind, we compute a lower limit in the expected scatter by assuming that the scatter between the masses is un-correlated. We add in quadrature the scatter of $\cML$  and $\MSHM$ from simulations (\RcMLAllScatter\ and \RMPSHMAllScatter, respectively, from \citealt{Remolina:20,Remolina:21}) with a $5\%$ scatter in $\MDLM$ \citep[from][]{Raney:20b}. This results in an expected scatter of \EcMLAllscatter\ for $\cML$ and \EMSHMAllScatter\ for $\MSHM$ that passed the visual inspection. In both cases, we find that the overall scatter measured in this work (\cMLAllScatter\ and \MPSHMAllScatter) is larger than expected. The difference between these scatters highlights some of the limitations in the simulation used by \citet{Remolina:20,Remolina:21} to account for the full range of scatter due to, e.g., baryonic effects, uncorrelated mass along the line of sight, and shear from nearby structures.


\section{Summary and Conclusions} 
\label{sec:conclusion}

A large number of strong lensing galaxy clusters is expected to be detected in current and upcoming large surveys. Estimating the mass at the core of these galaxy clusters will serve as one of the anchors to the radial mass distribution profile and measurement of the concentration. Detailed lens models to analyze these strong lensing clusters and measure the mass at the core of the galaxy cluster are limited by the small number of constraints available from the identified multiply imaged lensed sources and each can take multiple weeks to be finalized. Timely, efficient, and accurate methods to measure the mass at the cores of galaxy clusters in these large samples are needed. \citet{Remolina:20} assessed an empirically corrected mass $\cML$ enclosed by the lensing evidence, by assuming that their radial extent approximates the Einstein radius, and using the Einstein radius equation for a spherically symmetric lens. \citet{Remolina:21} assessed an aperture mass computed from single-halo lens models, $\MSHM$. Both papers  utilized simulated strong lensing images from the Outer Rim \citep{Heitmann:19}. In this work, we apply the two methods to observational data and use the publicly available detailed lens models from the SGAS, CLASH, HFF, and RELICS strong lensing cluster samples to evaluate the efficacy of the methods in measuring the core mass of galaxy clusters. We conclude the following:

\begin{itemize}
    \item The corrected mass enclosed by the approximate Einstein radius, $\cML$, has an overall scatter of \cMLAllScatter\ and bias of \cMLAllBias\ compared to the detailed lens models. The bias is reduced if large radii ($\eER >20\farcs0$) are excluded. For $\eER \leq 20\farcs0$ the scatter is \cMLSmallERScatter\ and the bias is \cMLSmallERBias.
    \item The SHM aperture mass when computed over the entire sample, has an overall scatter of \MSHMAllScatter\ and bias of \MSHMAllBias\ compared to the DLM. A quick visual inspection of the SHM outputs eliminates the SHM that fail to reproduce the lensing configuration, reducing the  scatter to \MPSHMAllScatter\ and the bias to \MPSHMAllBias.We find that the quick visual inspection is beneficial in reducing the scatter and bias between $\MSHM$ and $\MDLM$, and identify lines of sight that would benefit from a more detailed analysis.
    \item We confirm that in the region where the empirical correction was calibrated, $\eER<20\farcs0$, there is nearly no bias in $\cML$. We find that the bias in $\cML$ increases towards higher $\eER$ (see \autoref{fig:bin_systematics}). This trend could possibly be addressed by extending the work of \citet{Remolina:20} to larger cluster-centric radii and lower magnification.
    \item We explore the bias and scatter of $\MSHM$ and find a small positive bias and no trend with respect to the SHM-derived effective Einstein radius, $\RSHM$, the deviation from circular symmetry, nor number of multiply-imaged background sources with spectroscopic redshifts (see \autoref{fig:bin_systematics}). We find a slight trend in the scatter of $\MSHM$ in panels D and E, where the scatter decreases with increasing $\RSHM$ (panel D) and decreasing $\epsilon$ (panel E).
    \item To compare the overall scatter from $\cML$ and $\MSHM$ to that of simulations, we need to take into account the uncertainty in the detailed lens models. While we expect correlations between all mass estimates ($\cML$, $\MSHM$, and $\MDLM$), computing this is out of the scope of this analysis. We choose to compute a lower limit for the expected scatter by adding in quadrature $5\%$, which corresponds to the scatter of the mass from the detailed lens models, to the scatter measured in the simulation of \RcMLAllScatter\ for $\cML$ and \RMPSHMAllScatter\ for $\MSHM$ that passed the visual inspection. The resulting expected scatter is \EcMLAllscatter\ for the corrected mass enclosed by the approximate Einstein radius and \EMSHMAllScatter\ for the SMH that passed the visual inspection. The measured scatter in this work for both cases, \cMLAllScatter\ in $\cML$ and \MPSHMAllScatter\ in $\MSHM$, is higher than our estimated lower limit of the expected scatter. The difference is attributed to limitations in the simulation used by \citet{Remolina:20,Remolina:21} including baryonic effects, line-of-sight structure, and shear due to nearby structures.
    \item Detailed lens models are considered to be the state of the art in measuring the enclosed projected mass density within the cores of galaxy clusters. While likely underestimated, the relative \textit{statistical} lens modeling uncertainty of detailed lens models, marginalized over the large sample we investigated here, is of order \DLMstatsig. Systematic uncertainties are estimated in the literature \citep[e.g.,][]{Meneghetti:17,Raney:20b} at the $5-10\%$ level.  We show that the precision toll of using the significantly faster mass estimate methods is only a \MPSHMAllScatter\ or \cMLAllScatter\ increase over the detailed lens models. We conclude that if other, larger, sources of error dominate the analysis, these fast and efficient mass estimate methods become a powerful tool in analyses of large cluster samples.
\end{itemize}

Overall, this work demonstrates the successful application of these efficient methods to observational data as currently established, as well as their reliability to estimate the mass at the core of strong gravitational lensing galaxy clusters. We look forward to improvements to these methods benefiting from identification of strong lensing evidence by convolutional neural networks (e.g., \citealt{Canameras:20, Huang:21, Morgan:21}) and other machine learning algorithms to model the mass distribution of the SL clusters (e.g., \citealt{Bom:19, Pearson:19}).


\section*{Acknowledgements}
The authors would like to thank the anonymous referee for insightful suggestions that improved this manuscript.
We thank the HFF, RELICS, CLASH, and SGAS projects for making their lens models publicly available. Some of the High Level Science Products (HLSP) presented in this paper were obtained from the Mikulski Archive for Space Telescopes (MAST). STScI is operated by the Association of Universities for Research in Astronomy, Inc., under NASA contract NAS5-26555.
JDRG acknowledges support by the National Science Foundation Graduate Research Fellowship Program under Grant No. DGE 1256260. GM received funding from the European Union’s Horizon 2020 research and innovation programme under the Marie Skłodowska-Curie grant agreement No MARACAS - DLV-896778. Argonne National Laboratory’s work was supported by the U.S. Department of Energy, Office of Science, Office of High Energy Physics, under contract DE-AC02- 06CH11357.


\bibliographystyle{yahapj}
\bibliography{bibfile}

\end{document}